\documentclass[12pt]{article}

\usepackage[english]{babel}

\usepackage[letterpaper,top=2cm,bottom=2cm,left=3cm,right=3cm,marginparwidth=1.75cm]{geometry}

\usepackage{amsmath}
\usepackage{graphicx}
\usepackage[colorlinks=true, allcolors=blue]{hyperref}
\usepackage{rotating}

\title{Financial intermediation and risk in decentralized lending protocols}
\author{Carlos Castro-Iragorri\footnote{Corresponding author: carlos.castro@urosario.edu.co}, Julian Ramirez and Sebastian Velez\\
Universidad del Rosario}

\begin{document}
\maketitle

\begin{abstract}
We provide an overview of decentralized protocols like Compound and Aave that offer collateralized loans for cryptoasset investors. Compound and Aave are two of the most important application in the decentralized finance (DeFi) ecosystem. Using publicly available information on rates, supply and borrow activity, and accounts we analyze different elements of the protocols. In particular, we estimate ex-post margins that give a comprehensive account of the cost of financial intermediation. We find that ex-post margins considering all markets are $1\%$ and lower for stablecoin markets. In addition, we estimate quarterly indicators regarding solvency, asset quality, earnings and market risk similar to the ones used in traditional banking. This provides a first look at the use of these metrics and a comparison between the similarities and challenges to our understanding of financial intermediation in these protocols based on tools used for traditional banking.  
\end{abstract}



\section{Introduction}
\label{sec:intro}

Decentralized finance (DeFi) is the term used to describe decentralized applications or protocols running on a blockchain network (e.i. Ethereum) whose purpose is to provide financial services to cryptoasset investors. 
There is a fair amount of literature specially in computer science and more recently in finance that maps the different financial services offered in this ecosystem (\cite{Harvey2021}; \cite{Werner2021}; \cite{Schaer2021}; \cite{Jensen2021}; \cite{Chen2020}; \cite{Cai2018}). The different protocols are matched to existing services in the financial industry (exchanges, lending, derivatives, asset management, insurance) and new services like tokenization. Some of the benefits of DeFi with respect to the traditional financial industry have been systematically identified: efficiency, accessibility, open auditable, permisionless, interoperability, composability and non-custodial (\cite{Harvey2021}; \cite{Werner2021}; \cite{Schaer2021}).\\
Interest in DeFi is also due to observed and potential growth measured in terms of the value of crytoassets committed to the protocols: total value locked (TVL). In the Ethreum blockchain, where the most active DeFi protocols have been deployed, TVL went from $351.3$ million dollars in January 2020 to $54.13$ billion dollars in June of 2021. \\
Borrowing activity using cryptoassets can be done within the centralized exchanges, where users post assets as collateral to obtain loans that could be used to increase their positions in the same cryptoassets or others. This is in the same way as a brokerage account works where an investor with a significant amount of accepted securities has access to a line of credit. Lending can also be 
peer-to-peer (no intermediary required) meaning that a cryptoasset holder has to find another holder willing to lend his assets and set up some conditions under which they agree. The main innovation that DeFi lending protocols provide are pools of loanable resources that are automated through the use of smart contracts. All of the loans are collateralized and there is experimentation going on on non-collateralized loans. The most important lending protocols are Aave and Compound; these provide different markets to deposit accepted cryptoassets and use this collateral to obtain loans. On the other hand Maker (also classified as a lending protocol) provides a service to generate collateralized debt positions and its main use is to support a crypto-collateralized stablecoin, pegged to USD known as DAI. Aave and Compound at the end of June of 2021 have TVL values of $7.9$ and $7.2$ billion dollars, respectively. 
\\
Even though one of the benefits of the activity in DeFi is that the information is publicly available (open auditable), there are very few studies that provide an overview of the protocols based on the information. Most of the cited studies are based on characterizing and comparing the ideas behind the business models proposed in the white papers.  Having access to real-time and historical information on the blockchain is relevant because is an advantage compared to the problems of opacity and timeliness observed in the traditional financial industry.
The objective of this paper is to understand how financial intermediation is taking place inside of the DeFi lending protocols Compound and Aave. To understand the possible impacts of DeFi on the financial service industry it is relevant to compare the type of intermediation activities and the risk involved in light of what traditionally is considers as financial intermediation. In addition, using the data as opposed to the proposed business models can clarify some misconceptions around the protocols and help practitioners, regulators and academics understand the contribution and challenges of this technological innovation.\\
Some elements of protocols for loanable funds are analyzed in \cite{Gudgeon2020}. In particular, the authors provide an overview of the different interest rate models used in the protocols. In addition, they provide data on the historical behavior of interest and utilization rates and test efficiency in the markets by testing for uncovered interest rate parity and interdependence across the most important markets.   
Their results indicate that there are arbitrage opportunities in most of Compound's markets. It is not surprising then that there are other DeFi applications (e.g. yearn.finance, Curve Finance) that provide opportunities to optimally swap positions across markets to chase better returns.  One important role in terms of risk management provided by liquidators, acting on the protocol's user accounts has also been empirically analyzed by  \cite{Perez2020}  and \cite{Qin2021}.   
Our paper is related to literature on measuring financial intermediation ( \cite{Calice2018};\cite{Philippon2015}) and the role of technology in the financial service industry ( \cite{Thakor2020};\cite{Boot2021}). Quantifying financial intermediation in banks has been a tool that has been used for a long time to analyze economic development and efficiency within the industry. Most results indicate that the cost of financial intermediation is lower in more developed economies, these results have also been persistent over time. \cite{Philippon2015} offers a more comprehensive measure of the cost of intermediation in the financial industry in the US that accounts for more than just banks, a historical outlook (1886-2012) and assets. The quantitative results indicate that the annual cost for the US economy of financial intermediation has been stable and between $1.5\%$ to $2\%$. The stability over time also indicates that technological improvements do not appear to significantly decrease the unit cost of intermediation. One possible explanation for this puzzle discussed and provided by \cite{Gennaioli2015} is that technological innovation that does not increase trust among participants (investors and intermediaries) limit higher returns due to lower risk tolerance from investors. Because of fintech, understanding the effects of technological change in the financial service industry has been renewed. \cite{Boot2021} discussed some of the challenges that banks are starting to face from Bigtech platforms and opportunities from cooperating with specialized service providers. They mention the current strong points for banks such as relationship lending and access to stable funding. However, these two elements are some of the innovations we are seen in the decentralized lending space; where lending protocols have bootstrapped pools of resources from cryptoasset holders and through collateralized lending provide exchanges between unknown counterparties. \\
The introduction of these protocols in the financial service industry creates many challenges and questions: How can we start to measure the benefits of this new offer of financial services in the crypto space? for example, in terms of the cost of financial intermediation. What data and indicators should be used to give an overview of the financial soundness of these protocols? How can we compare these new financial services to traditional intermediaries? As far as we know this is the first paper that uses data on the protocols to measure the cost of financial intermediation and look at risk metrics for the Compound protocol. \\  
Our main findings show that the design of interest models does indeed provide larger ex-post margins for cryptoassets like ETH and WBTC than for stablecoins (DAI, USDC, USDT). The average margins for ETH and WTC are $1.37\%$ in Compound and $0.95\%$ in Aave, while for stablecoins these margins are $0.06\%$ and $0.29\%$, respectively. For Compound the overall margin taking into account all the income, expenses, borrow and supply activity from all of the markets is equal to $1\%$. This is lower than ex-post margins in traditional banking. It is important to note that this difference is affected by the fact that in DeFi all loans are collateralized while in banking they are mostly not collateralized. It is hard to say if the discount on collateralized lending is adequate or not, so a straightforward comparison between DeFi and bank lending is difficult.  
Although there have been periods of negative ex-post margins (in USDC and USDT) most of the time these margins are positive. However, what is a source of concern is that when we look at income from borrows minus the promised interest to be paid to suppliers, the operating margin is negative in the Compound protocol. Hence, you get a negative value for ROA and ROE with the current data. This poses a question regarding how are these protocols financing working capital and in particular, what is the expected return for investors. Like many other protocols in blockchain they obtain resources from initial coin offering and continuous minting of the protocols token COMP and AAVE. But it is expected that at some point there is some margin on intermediation services. The resources that are readily available in the protocols are from the reserve fund that takes a fraction of the interest income. However, these resources are strongly (in Compound) or weekly (in Aave) committed to mitigating some risks of the protocol.  We confirm findings that the level of undercollateralized positions are manageable and this means that the incentive structure for liquidators is adequate (there is room to adjust the close factor and avoid accumulating micro undercollateralized positions). Finally, since most of the collateral is posted in ETH and WBTC because of the historical volatility of these assets extreme scenarios of collateral shortfall should be considered and probably should adjust the level of reserves of the protocols. Currently, reserves only increase due to the borrowing activity of the protocol but do not take into account the potential magnitude of the risk they claim to cover.   
\\
The paper is structured as follows. Section \ref{sec:bmodel} gives an overview of how the protocols provide collateralized loans, the participants, risk management practices and provide historical information on the size of the most important markets. Section \ref{sec:data} explains how and where the publicly available information on the protocols can be obtained. Section \ref{sec:margins} discusses the interest rate model, provides historical information on the marginal supply and borrow rates, utilization and quantifies active, passive rates and intermediation margins in the most important markets.
Sections \ref{sec:risk} proposes a simple methodology to quantify some financial risks in the protocol using some elements of the CAMELS framework used in banks. Section \ref{sec:conclu} concludes.

\section{Business model}
\label{sec:bmodel}
Lending protocols like Compound and Aave are some of the most visible and important, measured by total value locked, applications in the decentralized finance ecosystem. As stated in the Compound whitepaper the objective of the protocol is to offer an automated money market for cryptoassets. There are two significant innovations of the protocol. The first is to provide collateralized loans using a pool of resources, where each pool determines a market. The borrowers obtain resources from the pool without any interaction with specific lenders.
The second innovation is the use of native token (Ctoken or Atoken) in each market that provides a unit of account and a mechanism to quantify and distribute the accrued interest to the lenders\footnote{A token can also be used to distribute losses among participants, for example in the case of insurance providers.}. \\
Loans are overcollateralized because the participants are anonymous, only identified by the blockchain address. Upfront collateral mitigates counterparty risk in this anonymous system but in addition, over-collateralization provides a buffer to account for the volatile value of the collateral. The cryptoassets accepted as collateral determine the market where the collateral can be posted and also the level of collateralization. The level of collateralization is a parameter of the protocol and according to the description of the protocol and discussions in changes to the parameters, it is intended to reflect the differences in market capitalization and price volatility among cryptoassets. 
For example, ETH has a collateral factor of $75\%$ meaning that if the user wants to borrow $3$ ETH then the minimum amount of collateral to support the loan is $4$ ETH\footnote{Collateral factor is the denomination in Compound, in Aave this parameter is known as the Loan to Value ratio (LTV).}. The inverse of the collateral factor is known as the collateral ratio, for ETH this critical ratio is $1.33$. This collateral reserve ratio is analogous to the required reserve ratio in fractional reserve banking, however, in the former, the collateral ratio is below $100\%$ meaning that few deposits can support a large portfolio of loans; this gives rise to the multiplier effect and creation of money in traditional banking.\\
The observed collateral ratio of an account in a protocol is an important health indicator in the collateralized loan markets, in particular Compound,
\begin{equation*}
    h_{i,t}=\frac{\text{Collateral amount}_{i,t}}{\text{Borrow amount}_{i,t}}
\label{eq:health}    
\end{equation*}
where $h_{i,t}$ is the health of account $i$ at time $t$\footnote{Time in the blockchain is determined by the block height, the block number. There is a timestamp in the block height indicating the date and time at which the block is mined. The time between blocks is not constant so is important to keep in mind that the time intervals are not regularly spaced.}. A loan in the protocol may be backed by different types of accepted collateral, therefore a loan in one market may be supported by the collateral in other markets, the important issue is that the collateral ratio of the portfolio is above the minimum required. 
As we will show in the next section (\ref{sec:participants}) these protocols were created to attract collateral in the form of cryptoassets like ETH and BTC and obtain loans in stablecoins. As expected the health indicator may deteriorate because the collateral loses value to the loans. This can happen because of shocks to the prices of the collateral but also because the stock of borrows includes the interest accrued on the loans and this means that, without additional resources posted as collateral, the rate of growth of the value of loans is greater than the rate of growth of the value of the collateral. It is important to recall that the collateral is deposited funds that earn a rate of return but to have a positive margin the protocol is set up so that the borrowing rate is strictly greater than the supply rate (see section \ref{sec:margins}).

\subsection{Participants}
\label{sec:participants}
\begin{figure}
\begin{center}
\includegraphics[scale=1]{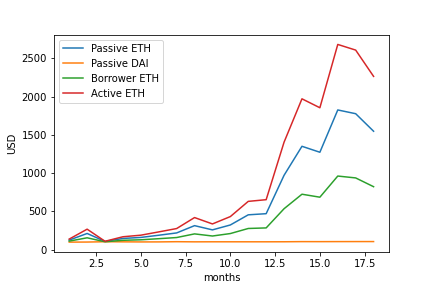}
\caption{The figure shows the value of net investment (sometimes including borrows from the protocol) of 100 USD in January 2020 with an investment horizon of one month and up to eighteen months (June 2021). The simulation is based on the historical daily data from the Compound protocol and considers different types of investors}
\label{fig:simulator}
\end{center}
\end{figure}

Four participants are interacting with the protocol: suppliers, borrowers, liquidators and protocol sponsors or distributed governance.\\
The distinction between suppliers and borrowers is arbitrary because of the collateral requirement every borrower must also be a supplier\footnote{The Aave protocol is exploring some form of non-collateralized loans where the borrower (obtains a credit line) is backed by a supplier that has posted collateral on his behalf.} The excess resources in a market (liquidity) are there because most users that supply collateral do not use their credit lines. These passive suppliers are interested in interest on their deposits and any additional benefits awarded in terms of distribution of the protocols own (governance) token. These suppliers can withdraw their cryptoassets at any time.
Suppliers enter the market by supplying and locking cryptoassets in the supply pool of a market (e.g. the \hyperlink{https://compound.finance/markets/ETH}{Compound ETH market} or \hyperlink{https://app.aave.com/reserve-overview/ETH-0xc02aaa39b223fe8d0a0e5c4f27ead9083c756cc20xb53c1a33016b2dc2ff3653530bff1848a515c8c5}{Aave ETH market})\footnote{Some digital wallets provide a direct link to the lending protocols.}. The number of resources supplied will allow the user to mint an amount of Ctokens determined by the current block exchange rate (e.g. $X_{t}^{ETH\mid cETH}$). An active supplier or borrower may now use these resources to borrow in the market where he supplied collateral or in any of the other markets. Supplied or borrowed funds do not have a term date, in the case of borrows the loan is maintained as long at there is more than enough collateral to support it. The comptroller is a risk management layer (a smart contract)  that approves or denies any additional transactions from the user. For example, a borrower cannot increase his borrows if there is not enough collateral. This is similar to obtaining a loan from a broker and using as collateral the investments (stocks, bonds or cash) that the user has in his broker account. The interest earned on deposits and paid on loans is time-varying and determined by the interest rate model in each market (see section \ref{sec:margins}).\\
If a depositor wants to exit the market he redeems (burns) the Ctokens for the underlying collateral at the current exchange rate. The value of this exchange rate will reflect the interest accrued during the time the collateral was held in the pool\footnote{Aave has a fixed exchange rate between the collateral and the Atokens and the interest accrued is paid in the form of additional tokens in the user's account.}.\\    
Borrowing resources in each market exposes users to different market risks. For example, if the user borrows an amount of ETH to be used outside of the protocol in USD he faces the uncertainty of the price of ETH in a later day that he plans to repay the borrows. On the other hand, a user may choose to borrow in a stablecoin market (USDT, USDC or DAI) to avoid the price volatility on the repayment value of the loan. In figure \ref{fig:simulator} we show the results of an investment simulator based on the historical data of the Coumpound ETH and DAI markets. The simulation considers an initial investment of 100 USD in January 2020. We consider passive investors as suppliers of the ETH or DAI market. At the end of the 18 months: The ETH passive investor has $1,546$ USD but this is mainly from the price appreciation of ETH, interest income is only $2.64$ USD. The DAI has $107.5$ USD mainly from interest income of $7.3$ USD. The ETH borrower supplies the same $100$ USD but takes out a loan of $50$ USD; at the end of the 18 months he still receives $1,546$ but the value of his initial loan is $773.4$ USD so his net position is $823.4$ USD. Finally, the active ETH investor supplies an initial amount of $100$ USD in the ETH market and take out a loan of $50$ USD in the DAI market and reinvest this amount on the ETH market, at the end of the 18 months his levered investment payoff $2,262.5$ USD. 
This simple simulation illustrates that the optimal strategy for active investors in these protocols is to increase their bets on the price of cryptoassets used as collateral by borrowing in stablecoins. However, they need to keep a buffer to that price volatility of the cryptoassets like ETH and BTC do not put them in a situation of forced liquidation of their positions.  
\\
As we mentioned previously the collateralized loans are similar to those obtained from a broker. However, because of the automation feature of smart contracts, there is no room for margin calls. The protocol has to deal with the risk of systematic under collateralization of account to the extent that such a situation undermines the credibility of the suppliers to the protocol. The solution to this problem is to create incentives for a liquidator to monitor the health \ref{eq:health} of accounts that borrow resources. This innovation allows the protocol to outsource the oversight of the accounts that are borrowing funds from one or more of the markets.\footnote{This innovation in the form of a seeker was introduced by the Defi protocol, Maker Dao, known for introducing a crypto collateralized stablecoin pegged to USD.}
Liquidators are constantly monitoring the health of the account looking accounts with health indicator below 1 because these accounts are subject to liquidation\footnote{Although the health indicator is mentioned in the Aave whitepaper liquidators monitor if the ratio of the loan value to the collateral value of a borrower falls below a market defined parameter known as the liquidation threshold.}. A liquidator calls a function within the smart contract to liquidate a particular borrower. The liquidator may transfer the borrowed amount, however, he will not be able to repay the entire loan, only a fraction of the loan given by a close factor. The close factor is another parameter of the protocol set to $50\%$ this means that the liquidator can only repay half of the value of the loan. After this repayment, the account's health indicator will improve sufficiently to avoid liquidation, if not the same liquidator or another will get the opportunity to repay half of the outstanding amount. Liquidation is costly because the liquidator must pay the transaction fee requires to execute the smart contract. To incentive liquidators, there is a liquidation incentive or penalty, another parameter of the protocol, that will increase by $1.08\%$ the value of the token received\footnote{In Compound is known as a liquidation incentive for the liquidator while in Aave it is known as a liquidation penalty because the borrower has to pay for it. Even though the denomination is different in the protocols this parameter determines the gain for liquidators.}. For example, a liquidator liquidates an outstanding loan of 4 units of token ZRX, then the maximum to liquidate is 2 units of ZRX because of the close factor. The gross benefit is the liquidation incentive, the difference between the paid out loan and the value received $2*(1+1.08\%)-2=0.0216 ZRX$. The amount paid to the liquidator $2.0216$ is seized from the collateral of the account that was liquidated. This creates an incentive for borrowers to have enough collateral, a buffer above the threshold, to avoid being liquidated. The net benefit to the liquidator must take into account the transaction cost of executing the function of the smart contract and any additional trading fee to exchange the token received to the liquidators preferred currency or asset. This creates a risk for the protocol because not all the accounts that need to be liquidated will be effectively liquidated because the cost of liquidation will be greater than the net benefit. This is a risk of great concern for the protocol.  That there is a shock to the price of the cryptoassets used as collateral that will increase the demand for liquidators, but a the same time the cost of liquidation increases because of congestion in the Ethereum network creating a systemic problem of undercollateralized accounts.                    
\\
These types of protocols have different elements: design decisions and parameters. When the protocol is created and transitions from a testing environment to a live product these elements are defined by the protocol sponsors. Some of these design decisions answer the following questions: what crytoassets are accepted as collateral? what interest rate model is used in each market? how to make adjustments to the protocol to save on transactions fees? On the other hand, there are specific parameters on the protocol and the markets: collateral factor, reserve factor, liquidation incentive and close factor.\\
At some point, the protocol sponsors extend the design and parameter decisions to the community of users. This decentralized community of holders of the protocol's token vote on a proposal to upgrade the protocol. Where the proposals and the votes are publicly available information. protocol sponsors among which there are initial investors have initially a large stake in terms of the votes but lead the proposals. 

\subsection{Monetary aggregates in the markets}
\label{sec:mkts}
A decentralized lending market for a particular cryptoasset is a self-sustainable money market. The pool of resources in each market represents the monetary aggregates.
The monetary aggregates use as unit of account (measurement of value) the cryptoassets\footnote{This means that all relevant monetary aggregates are in units of the cryptoasset accepted as collateral, except when the exchange rate is updated.} and must comply with the following monetary identity: 
\begin{equation}
	TS_{t} = C_{t} + TB_{t} - R_{t}
	\label{eq:identityM}
\end{equation}
where $TS_{t}$ is the total supply,$C_t$ is cash, $TB_{t}$ is the total borrows and $R_t$ is reserves. Cash is a construct that reflects the disposable liquidity, however, the available market liquidity $MktL_{t}$ or liquidity pool excludes the reserves. Therefore,
\begin{equation}
	MktL_{t} = TS_{t} - TB_{t}, 
	\label{eq:liquidity}
\end{equation}
and hence $C_{t} = MktL_{t} + R_{t}$. The most important indicator for market liquidity in the protocols is the utilization rate,
\begin{equation*}
    u_{t}=\frac{TB_{t}}{TS_{t}}.
\label{eq:utilization}
\end{equation*}
As we will see in the next section \ref{sec:margins} the utilization rate determines the interest rate on loans. For higher utilization rates (more borrows to the total supply) of the protocols then the interest rate increase reflecting scarcity.\\
In Compound, the objective of the reserves is to be used when liquidators cannot get into the undercollateralized loan fast enough or are not willing to liquidate a position\footnote{In the Aave protocol the reserve fund is intended to promote the growth and development of the protocol by paying the protocol sponsors.}. In both cases, this creates an undercollateralized position where the current collateral does not cover $100\%$ of the loan.  This fund is known as a reserve fund for the risk associated with undercollateralization. Reserves will accumulate as a function of the accrued interest in every block for a given market. In other words, reserves in market $j$ will be increased due to the loans and the interest accrued by the borrowers,  
\begin{equation}
	R_{t,j} = R_{t-1,j} + TB_{t-1, j} \cdot \Delta \cdot
	\frac{b_{t-1, j}}{blocksPerYear} \cdot reserveFactor_{j}
	\label{eq:reserveSolvency}
\end{equation}
where $reserveFactor$ is a parameter defined by governance for each collateral cryptoasset. The effective accumulation rate is blockrates meaning that the borrowing rate $b_{t-1,j}$ (that we discuss in the next section \ref{sec:margins}) are divided by the expected number of blocks per year. The number of blocks per year is estimated based on the assumption of the seconds it takes to mine a block $13.15$ seconds; this implies approximately 2.4 million blocks per year. In addition to the reserve fund, the Aave protocol has implemented an insurance pool known as the safety module where investors buy the protocol token and stake their investment on a pool to cover losses from systemically undercollateralized positions. The investors receive periodical rewards in additional units of the token but must cover the possible losses.\\
Reserves are accumulated base on the stock of loans in each market and hence are no estimated based on the amount of risk. This approach deviates from current risk management standards where such reserve funds (e.g. in central counterparty clearing houses) are estimated based on the potential exposure under normal and stressed scenarios. This implies that the value of the reserved fund in these markets could be lower than the risk that they claim to potentially mitigate. It is an open question to determine the level of underestimation of the reserve fund\footnote{A methodology is required to map the risk of undercollateralization to the scale of the market and its sensibility to shocks. }.\\  

\begin{figure}[ht] 
  \label{ fig7} 
  \begin{minipage}[b]{0.5\linewidth}
    \centering
    \includegraphics[width=1\linewidth]{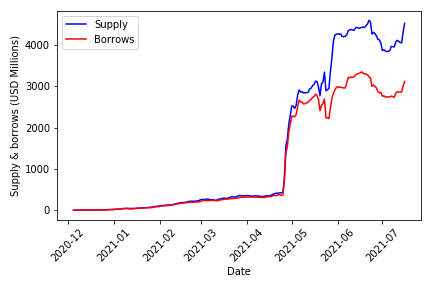} 
    \caption{\scriptsize USDC Total Borrow and Supply in USD Millions Aave}
    \label{fig:usdcAggAave}
    \vspace{4ex}
  \end{minipage}%%
  \begin{minipage}[b]{0.5\linewidth}
    \centering
    \includegraphics[width=1\linewidth]{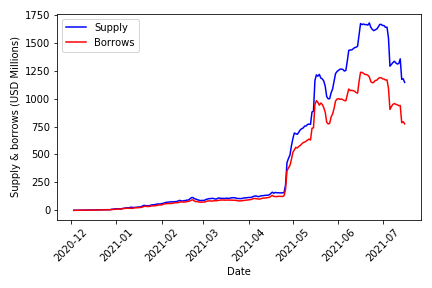} 
    \caption{\scriptsize DAI Total Borrow and Supply in USD Millions Aave} 
    \label{fig:daiAggAave}
    \vspace{4ex}
  \end{minipage} 
  \begin{minipage}[b]{0.5\linewidth}
    \centering
    \includegraphics[width=1\linewidth]{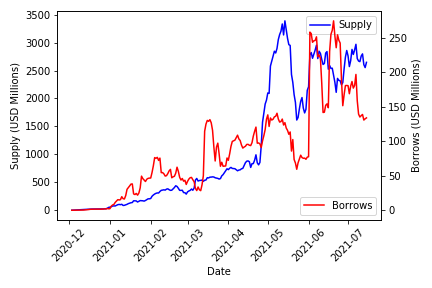} 
    \caption{\scriptsize ETH Total Borrow and Supply in USD Millions Aave}
    \label{fig:ethAggAave}
    \vspace{4ex}
  \end{minipage}%% 
  \begin{minipage}[b]{0.5\linewidth}
    \centering
    \includegraphics[width=1\linewidth]{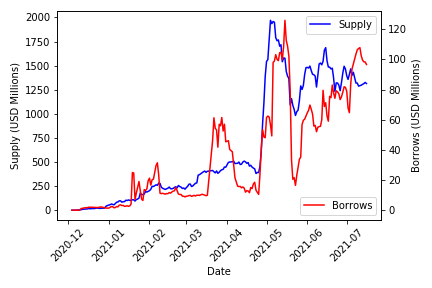} 
    \caption{\scriptsize WBTC Total Borrow and Supply in USD Millions Aave}
    \label{fig:wbtcAggAave}
    \vspace{4ex}
  \end{minipage} 
\end{figure}

Figures \ref{fig:usdcAggComp} through \ref{fig:wbtcAggAave} shows the aggregates of the most important markets: USDC, DAI, ETH and WBTC. In the short life span of the protocols Compound, one year and a half and Aave version 2, six months the total supply in USDC, DAI and ETH are around 2.5 to 5.5 billion dollars in each market and around 1 and 1.25 billion in WBTC. Total borrows are also large in stablecoins  1 to 4 billion depending on the specific market and protocol and 100 to 172 million in ETH or WBTC. As a point of comparison at the end of the second quarter of 2021 the amount of \hyperlink{https://fred.stlouisfed.org/series/DPSACBW027SBOG}{deposits in commercial banks in the US} was approximately $17,000$ billion dollars, and the amount of \hyperlink{https://fred.stlouisfed.org/series/H8B3094NCBAM}{borrows} was approximately $1,638$ billion dollars.    

\section{Data}
\label{sec:data}
The most important applications of decentralized finance are currently running in the Ethereum blockchain, these include the Compound and Aave protocol. Since the transactions are registered in this publicly accessible database is it possible to access the account information of the users and recover information for the different asset markets. These protocols provide a historical API service\footnote{The API information provided by the protocols is available in the following links: \href{https://compound.finance/docs/api}{Compound} and \href{https://aave-api-v2.aave.com}{Aave}.} to obtain historical information on the markets. We use the API's provided by the protocols to obtain the daily information on rates and additional information: total supply and borrows, cToken exchange rate and the USD price of the underlying cryptoasset.\\ 
More recently a project known as \href{https://thegraph.com/}{The Graph} has been developing a decentralized system to query the information contained in public blockchains. This protocol provides different subgraphs that provide information regarding some of the protocols in the Ethereum blockchain. The subgraphs provide dictionaries (schemas) that allow users to write queries base on the GraphQL query language to get information regarding the markets, accounts and transactions. We use python and the API service to query the information on markets and accounts for particular block heights in the Ethereum blockchain. From the queried accounts we recover information on the interest accrued on the borrow and supply side, the total amount borrowed and/or supplied and the amount of collateral. 
%Each blockheight has a timestamp therefore the information at that point in time represents        

\section{Interest rate models and inter mediation margins}
\label{sec:margins}
\begin{figure}[ht] 
  \label{ fig7} 
  \begin{minipage}[b]{0.5\linewidth}
    \centering
    \includegraphics[width=1\linewidth]{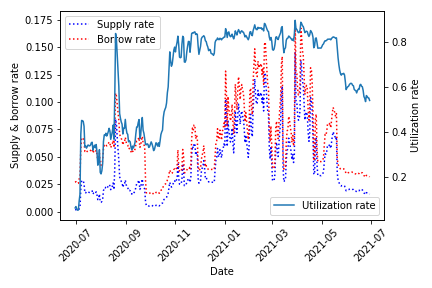} \caption{\scriptsize USDC rates Compound}
    \label{fig:usdcRatesComp}
    \vspace{4ex}
  \end{minipage}%%
  \begin{minipage}[b]{0.5\linewidth}
    \centering
    \includegraphics[width=1\linewidth]{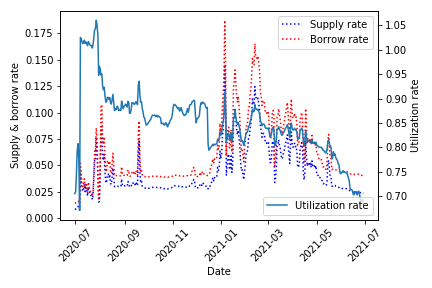} 
    \caption{\scriptsize DAI rates Compound} 
    \label{fig:daiRatesComp}
    \vspace{4ex}
  \end{minipage} 
  \begin{minipage}[b]{0.5\linewidth}
    \centering
    \includegraphics[width=1\linewidth]{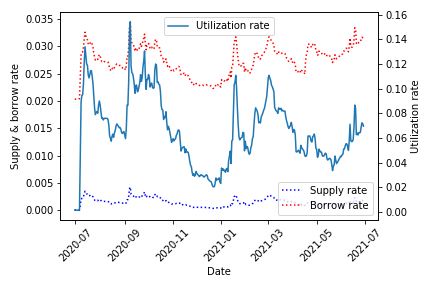} 
    \caption{\scriptsize ETH rates Compound}
    \label{fig:ethRatesComp}
    \vspace{4ex}
  \end{minipage}%% 
  \begin{minipage}[b]{0.5\linewidth}
    \centering
    \includegraphics[width=1\linewidth]{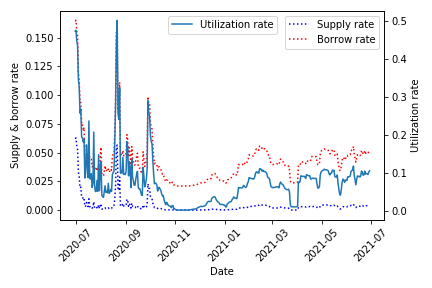} 
    \caption{\scriptsize WBTC rates Compound}
    \label{fig:wbtcRatesComp}
    \vspace{4ex}
  \end{minipage} 
\end{figure}

\begin{figure}[ht] 
  \label{ fig7} 
  \begin{minipage}[b]{0.5\linewidth}
    \centering
    \includegraphics[width=1\linewidth]{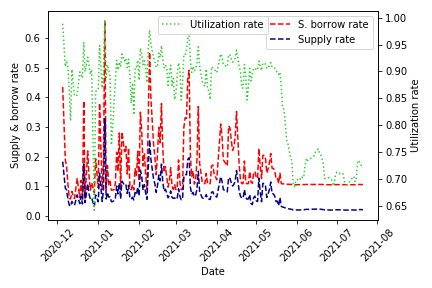} 
    \caption{\scriptsize USDC rates Aave}
    \label{fig:usdcRatesAave}
    \vspace{4ex}
  \end{minipage}%%
  \begin{minipage}[b]{0.5\linewidth}
    \centering
    \includegraphics[width=1\linewidth]{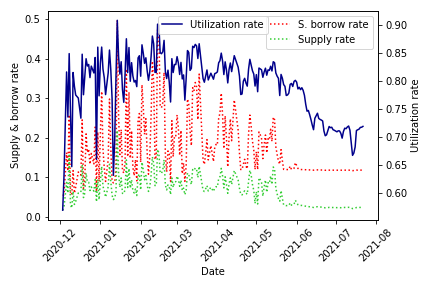} 
    \caption{\scriptsize DAI rates Aave} 
    \label{fig:daiRatesAave}
    \vspace{4ex}
  \end{minipage} 
  \begin{minipage}[b]{0.5\linewidth}
    \centering
    \includegraphics[width=1\linewidth]{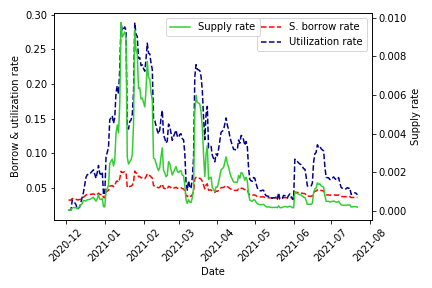} 
    \caption{\scriptsize ETH rates Aave}
    \label{fig:ethRatesAave}
    \vspace{4ex}
  \end{minipage}%% 
  \begin{minipage}[b]{0.5\linewidth}
    \centering
    \includegraphics[width=1\linewidth]{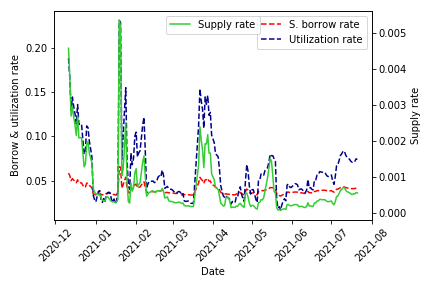} 
    \caption{\scriptsize WBTC rates Aave}
    \label{fig:wbtcRatesAave}
    \vspace{4ex}
  \end{minipage} 
\end{figure}

%where $t$ denotes block time
The borrow and supply for each market are determined by the current interest rate model.
Loans are granted with variable interest rates\footnote{Aave has introduces fixed (stable) rate loans. The rates are fixed in the short term but are subject to covenants: e.g. if the utilization rate is above $95\%$ the rate may be increased.}.
These models are coded into the smart contracts and their parameter can be adjusted based on the current governance of the protocol. \\
The rates quoted are the expected yearly rates. For example, in the market for ETH, the main component of the model is the functional form of the borrow rate,
\begin{equation}
    b_{t}=\alpha + \beta u_{t},
    \label{eq:borrowR}
\end{equation}
where $u_{t}$ is the utilization rate. Hence the borrow rate is a linear function of utilization. The gross supply rate is a function of the borrow rate,
\begin{equation}
    s_{t}^{g}=b_{t}u_{t}=(\alpha + \beta u_{t})u_{t}=\alpha u_{t}+\beta u_{t}^{2}
    \label{eq:gsupplyR}
\end{equation}
The gross supply rate is a quadratic function of utilization. The net supply rate consider the part of the rate that is set aside for the reserve,
\begin{equation}
    s_{t}^{n}=s_{t}^{g}(1-\psi)=b_{t}u_{t}(1-\psi)
    \label{eq:nsupplyR}
\end{equation}
where $\psi$ $\in (0,1)$ is the reserve factor for that asset. The information on the current rates can be observed using the dashboard provided by the protocols, for example for the ETH market. %(links compound, aave,...).
\\
The quoted intermediation margin of the protocol,
\begin{align}
    m_{t}&=&b_t-s_{t}^{n}(1-\psi)=b_t(1-u_{t}(1-\psi))\\
    &=&\alpha+(\beta-\alpha(1-\psi))u_{t}-\beta(1-\psi)u_{t}^{2}.
    \label{eq:qmargin}
\end{align}
Because of the quadratic form of the supply rate, it is not difficult to see that the quoted margin is also a quadratic function of the utilization rate. Furthermore, from this function, it is possible to find an optimal value of the utilization that maximized the margin in the protocol,
\begin{equation*}
    u_{t}^{*}=\frac{1}{2}(\frac{1}{1-\psi}-\frac{\alpha}{\beta}).
\label{eq:optutil}    
\end{equation*}
The optimal utilization rate is a function of the parameters of the borrow rate and the reserve factor.\\ 
For stablecoins such as USDC, DAI and USDT the protocols use a kinked rate model that introduce a different slope as a function of a particular value of utilization $u^{kink}$  \cite{Gudgeon2020}. 
\begin{equation}
    b_t=\left\{\begin{array}{ll}
        \alpha + \beta_t u_t & \text{, if }u_t\leq u^{kink} \\
        \alpha + \beta u^{kink} + \gamma(u_t-u^{kink}) & \text{, if }u_t> u^{kink}
    \end{array}\right.
\label{eq:borrowRS}
\end{equation}
These kinked models are designed to avoid high utilization rates by significantly increasing the cost of borrows (slope of the model) when the utilization crossed the utilization threshold $u^{kink}$. High utilization rates (near $100\%$) created liquidity risk in the protocol because it creates frictions on the withdraw of funds from the protocol. Over the history of the protocol administrator modified and introduced proposals to adjust the parameter of the stablecoin markets in order to increase the cost of borrowing above the target utilization.\\
In these markets the quoted intermediate margin is,
\begin{equation}
\footnotesize
    m_t=\left\{\begin{array}{ll}
        \alpha+(\beta-\alpha(1-\psi))u_{t}-\beta(1-\psi)u_{t}^{2} & \text{, if }u_t\leq u^{kink} \\
        \alpha+(\beta+\gamma \psi) u^{kink} + (\gamma+(\alpha+\beta u^{kink})(1-\psi))u_{t}  - \gamma(1-\psi)u_{t}^{2} & \text{, if }u_t> u^{kink}
    \end{array}\right.
\label{eq:qmarginS}
\end{equation}
%Althought we can calculate an optimal utilization rate from the point of view of revenue maximization above $u^{kink}$ as: 
%\begin{equation*}
%    u_{t}^{*}=\frac{1}{2(1-\psi)}-\frac{\alpha+\beta u^{kink}}{2\gamma}.
%\label{eq:optutilS}    
%\end{equation*}
%it is important for the protocol to have utilization rates below $u^{kink}$ to avoid liquidity risk.
Figures \ref{fig:usdcRatesComp} thorough \ref{fig:wbtcRatesAave} shows the utilization, borrow, and supply rates for the most important markets in Compound and Aave protocols. We can see that utilization rates are systematically higher, above $70\%$, for the stablecoin markets than in the other asset markets, this is expected because as we mentioned in section \ref{sec:participants} there is less uncertainty regarding the expected amount to be repaid in fiat currency in these markets. Utilization in the ETH and WBTC markets is usually below $10\%$ and it is also affected by periods of price volatility in the assets that are mainly used for collateral.
%The figures \ref{fig:mratesC} and \ref{fig:mratesA} also indicate how far is the empirical utilization from the optimal utilization that maximizes profit for the protocol..... \\
Variable borrow rates in stablecoin markets are usually around $4\%$ with sharp increases during periods where utilization jumps. Aave offers stable borrow rates for USDC and DAI with rates closer to $10\%$.
The margin stablecoin markets is not larger because high utilization rates by design guarantee that supply rates are not too far behind borrow rates, on average the stated margin at the end of June 2021 is around $1.37\%$. On the other hand, the stated margin in ETH and WBTC in Compound is between $3\%$ and $3.7\%$, and in Aave, these are higher, $5.2\%$ and $8.9\%$ respectively.\\
As a point of comparison, centralized solutions like \hyperlink{https://blockfi.com/}{Blockfi} offers collateralized loans (accepting BTC and ETH as collateral) in USD at rates of $4.5\%$. Personal loans (non-collateralized) rates at commercial banks in the US, during the same period, are above $9\%$\footnote{\hyperlink{https://fred.stlouisfed.org/series/TERMCBPER24NS}{FRED: Finance Rate on Personal Loans at Commercial Banks, 24 Month Loan}}, average $15$-year fixed rate mortgage is around $2.3\%$ \footnote{\hyperlink{https://fred.stlouisfed.org/series/MORTGAGE15US}{FRED: 15-Year Fixed-Rate Mortgage Average in the United States}}, and collateralized loans at brokerages offer variable rates starting at $4\%$ plus 30-day LIBOR\footnote{\hyperlink{https://www.tdameritrade.com/investment-products/collateral-lending-program.html}{TD Ameritrade}}.\\
In the ETH and WBTC, borrow rates are around $2.5\%$ in Compound and $5\%$ in Aave. In these markets, the margin is much higher 
because the non-linearity in the interest rate models at low utilization rates reduce substantially the supply rate. 
\\
In the banking literature, the intermediation margins can be measured as the marginal rate based on the quoted rates of the different products offered by the financial institution. Active rates will be based on the different types of loans (e.g. consumer, commercial, mortgage) and the passive rates are based on the different products offered to depositors (saving and checking accounts, CDs). However, these marginal rates are only indicative of the most recent operations and do not take into account issues such as the quality of the loans. \\
To have a complete picture of the intermediation margins it is necessary to estimate the ex-post spreads. The ex-post
spread is the difference between banks' actual interest revenue and their actual interest expense expressed as a ratio of the underlying loans and deposits, respectively\footnote{If we are only interested in the operational margin then net income (interest revenue minus expenses) is divided by the total number of assets}.
The ex-post active, passive rates and spread give a more complete picture of the cost of financial intermediation and are frequently used to compare how different forms of financial frictions, inefficiencies, market power affect the development of financial markets around the world, \cite{DemirgucKunt1999}. 
In \cite{Calice2018} the authors provide a recent benchmark of intermediation margins around the world (160 countries, data from 2005 to 2014). The authors estimate the net interest margins for each of the countries in the sample. Some of the margins for groups of countries are $1.75\%$ for OECD high income ($2.8\%$ for US), $4.3\%$ Latin America and the Caribbean and $3.6\%$ South Asia.  The implications of the paper still show that lower-income countries face higher intermediation margins because of inefficiencies due to the large overhead cost of providing financial services. 
\\
These average intermediation margins are based on accounting information on the banks. This information is usually obtained from the banking regulators within countries\footnote{For cross country studies the usual source of this information is Bankscope.}.\\
Although decentralized lending protocols are not currently regulated and hence required to provide regular accounting information, the information regarding the transactions and account balances are stored in a blockchain. We can use the different API's services presented in section \ref{sec:data} to obtain the historical information for the behaviour of the interest rate model in each of the asset markets. In addition, we obtain quarterly information on the supplier and borrower accounts of each market to estimate the interest collected and paid or accounted for at a particular block height.\\

Tables \ref{tbl:exratesComp} and \ref{tbl:exratesAave} provides the ex-post rates estimated from the account information quarterly from June 2020 to June 2021 for the Compound protocol and December 2020 to June 2021 for the Aave protocol\footnote{For Aave the shorter sample is because the launch of the current version (known as version 2) of markets was at the end of 2020.}. We find a strong variation in the active, passive and margins in the sample and across markets. This is probably due to the protocols are relatively new and because the activity of the protocols is strongly related to the price of the underlying collateral like ETH and BTC. As expected and because of the design of the interest rate models, the average margin for the ETH and WBTC marked is higher ($1.37\%$ in Compound and $0.95\%$ in Aave) than for the stablecoins ($0.06\%$ in Compound and $0.29\%$ in Aave). If we consider all of the other markets for collateral in Compound (including cryptoassets like REP, BAT, ZRX, UNI, COMP, and LINK) then the average margin is $1\%$, this is lower by $0.75\%$ ($3.3\%$)  compared to OECD high-income countries (Latin America and the Caribbean) intermediation margins. It is also lower by $0.5-1\%$ to the alternative approaches of measuring the historical cost of financial intermediation in the US discussed in section \ref{sec:intro}. However, it is important to take into account that these protocols provide collateralized loans ad opposed to traditional financial intermediation where most of the loans are not collateralized. Banks rely heavily on the trust relationships that are built through continuous interaction between the institutions and their clients. One could argue that the reduced cost of financial intermediation provided by these protocols is the result of using existing interoperable infrastructure (e.g. Ethereum network and Dapps) and requiring a minimal amount of trust among participants. They also leverage the use of the native cryptocurrencies embedded in the incentive system of the blockchain and other tokens as collateral. Many open research questions remain as to how much of the benefits are due to the technology and/or the massification of collateralized lending based on digital or non-digital assets.

\section{Risk}
\label{sec:risk}
Banking is a highly regulated activity because unlike other enterprises the working capital is obtained from a large base of depositors that used financial services (payment system, investments among others).  In its short existence, decentralized finance has been successful at leveraging the benefits of blockchain technology: facilitating access to markets, interoperability across services, modularity and composability of services, transparency and distributed governance.
Some of these issues are well-recognized pain points of the legacy information systems that are the infrastructure that supports financial transactions, \cite{Harvey2021}. However, to compete in the traditional financial service industry it is necessary to have a similar supervisory oversight that provides a sufficient amount of protection on consumers while at the same time allowing for these types of innovations.\\
Banking supervision has come a long way; the Basel accords intended to provide a level playing field for institutions that were internationally active and hence try to avoid any form of regulatory arbitrage across jurisdictions. One of the most simple approaches to compare financial institutions is using the CAMELS rating system. The CAMEL rating system was introduced in the early 1980s in the US and has been consistently used by supervisors (the Federal Reserve, the FDIC, and the OCC) to perform periodic examinations and compare the health of institutions among each other and across time\footnote{The ECB also performs a similar indicator based review on banks, this exercise is known as Supervisory Review and Evaluation Process (SRVP).}. The evaluation based on the CAMELS indicators by the supervisors is shared with senior management and depending on the overall performance the supervisors may ask the institutions to correct the issues found in the evaluation. The most current version of the rating system considers six components: Capital adequacy, Asset quality, Management, Earnings, Liquidity risk and a bank's Sensitivity to market risk. Within each of these components, there are several indicators used to assess the bank's conditions along with this components. Then the indicators are aggregated to provide a synthetic rating scale from 1 "strong" to 5 "critically deficient". A well-known drawback of the CAMELS rating system is an ex-post measure so it needs to be complemented with more forward-looking risk management tools based on stress testing.\\

In decentralized finance, there are no agreed-upon measures on the performance of the business models or risk management issue on the existing protocols. However, this does not mean that protocols are not aware of the different risks that they are exposed to. \cite{Werner2021} provide a classification of risk (security) factors that are important for these protocols as well as documented cases of realizations of these risks. The classification considers technical risks (e.g. smart contract vulnerabilities, single transaction or transaction ordering attacks) that affect the security of deposited funds in the protocol. Economic risks are related to the design of the protocol. Mitigating economic risk requires the right incentive among the participants. Some of these mitigators mentioned by the authors have been discussed in the context of lending protocols in section \ref{sec:bmodel}: overcollateralization, market and oracle manipulation, governance risk, threats from Miner Extractable Value. Finally, the authors recognized that there are still many open questions regarding technical or economic risk.\\
Protocols in decentralized finance have also stepped up efforts to have independent external audits on technical and economic risk. \href{https://compound.finance/docs/security#audits}{Compound} and \href{https://docs.aave.com/developers/security-and-audits}{Aave}  protocols provide information on the companies that have performed audits on technical and/or economic risk.  For economic risk \cite{Gauntlet2020} provides a detailed market risk assessment of the protocol. The purpose of this technical document is to provide a framework for market risk stress testing of the markets in the protocol and in particular the risk of undercollateralization. As mentioned in section \ref{sec:bmodel} the most important risk on these collateralized lending protocols is that the value of the collateral is unable to sustain the value of the borrowed resources. The value of the collateral is susceptible to the volatility of the price of the crypto assets accepted as collateral.  
The framework is uniquely designed for decentralized finance since it uses a simulated agent-based model (the model is based on \cite{Chitra2019}) running on an environment that replicates the behaviour of the smart contracts. The report concludes that ETH market as it was parametrized at the time of evaluation (2019) could have a substantial increase in utilization via an increase in the demand for borrows while at the same time the risk of undercollateralization (of the outstanding borrows) would be less than $1\%$. The simulation of the price of ETH is based on historical information observed and scenarios with increased volatility. They also show that the liquidation incentive of $105\%$ will attract enough liquidators to liquidate most of the undercollateralized positions. \\
From a community perspective there is an important interest in developing frameworks to quantify risk in decentralized lending protocols. One project is \href{https://defiscore.io/overview}{Defi score} which intends to develop a simple score with a range from 1 to 10 to measure protocol risk along with some factors: smart contract risk, collateralization, governance, liquidity among others. In addition, there are specialized news outlets like \href{https://www.theblockcrypto.com/data/decentralized-finance/cryptocurrency-lending}{The Block} that provide a historical comparison of different protocols in terms of revenue, lending and value locked (supply). 
\\
In this section, we use some of the indicators that are part of the CAMELS framework for banks and apply them to the information obtained from these lending protocols. We explore how they can be adapted and understood in the context of economic risk analysis for decentralized finance. We are not able to consider all of the indicators, for example, management (operational expenses) because there is no publicly available financial information. We also exclude liquidity because this is captured by the utilization rate and it has been discussed in section \ref{sec:margins}.
As explained in section \ref{sec:data} we obtain information on the lending markets and the accounts from API service provided by the protocol and The Graph. The purpose of the exercise is not to derive a synthetic rating for each protocol but rather to estimate some components of the CAMELS rating system in the context of decentralized lending protocols and discuss the benefits and limitations.

\subsection{Capital adequacy}
To determine capital adequacy of a financial institution the usual measure is the solvency ratio,
\begin{equation*}
    \text{Solvency Ratio}=\frac{\text{Tier 1 K}+\text{Tier 2 K}}{\text{Risk weighted assets}}.
\end{equation*}
The distinction between Tier 1 capital and Tier 2 capital, is that the former is readily available before default. In section \ref{sec:mkts} we mentioned that in each market $j$ has a reserve fund, where $R_{t,j}$ denotes the value of this fund at time $t$. One of the objectives of the fund is to cover any losses resulting in systemically undercollateralized accounts. The assets in each market are the loans granted, but we need to weigh each of the loans by a risk weight. We propose the weights to be defined base on risk buckets determined by the value of the health indicator of the account, \ref{eq:health}. Table \ref{tbl:riskBuckets} present the five risk buckets, where higher weights are given to riskier loans. As in the 1988 Basel accord, the risk-weighted assets are each of the loans multiplied by its corresponding weights. In table \ref{tbl:camels} we present the value of the reserve fund which has increased from less than one million dollars to $27.3$ million in over a year. We use the risk buckets to weigh the outstanding borrows taking into account the health of each account and obtain the risk-weighted assets (RWA). The reserves and the RWA are for all of the markets because collateral can be used to obtain loans in different markets.
The solvency ratio is less than $1\%$ but has been increasing as borrowing activity increases. In US banks the solvency ratio has been steadily increasing since and currently is close to $12\%$. Under such standards, the protocols are undercapitalized, however, one has to take into account that these loans are collateralized as opposed to most bank lending. The level of reserves in the protocol increases when the interest is accrued based on the loan and the reserve factor of each market. The drawback of this approach is that this mechanism to build up reserves does not take into account the potential risk of undercollateralization. This is important because the value of collateral can change substantially given the volatility of crypto assets like ETH and BTC. 

\subsection{Asset Quality}
\begin{figure}[ht] 
  \label{ fig7} 
  \begin{minipage}[b]{0.5\linewidth}
    \centering
    \includegraphics[width=1\linewidth]{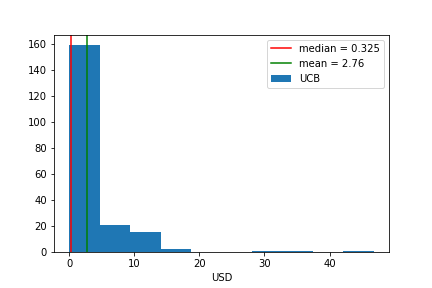} 
    \caption{\scriptsize Distribution of UCB, II-20}
    \label{fig:hUCBQ220}
    \vspace{4ex}
  \end{minipage}%%
  \begin{minipage}[b]{0.5\linewidth}
    \centering
    \includegraphics[width=1\linewidth]{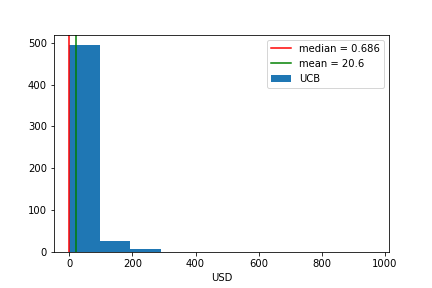} 
    \caption{\scriptsize Distribution of UCB, IV-20} 
    \label{fig:hUCBQ420}
    \vspace{4ex}
  \end{minipage} 
  \begin{minipage}[b]{0.5\linewidth}
    \centering
    \includegraphics[width=1\linewidth]{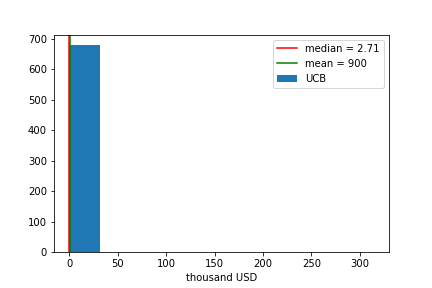} 
    \caption{\scriptsize Distribution of UCB, I-21}
    \label{fig:hUCBQ121}
    \vspace{4ex}
  \end{minipage}%% 
  \begin{minipage}[b]{0.5\linewidth}
    \centering
    \includegraphics[width=1\linewidth]{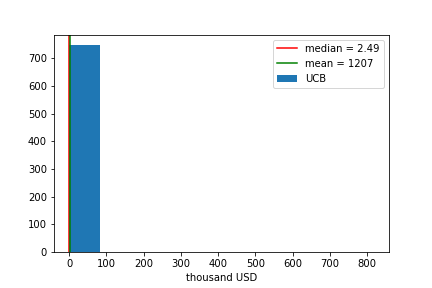} 
    \caption{\scriptsize Distribution of UCB, II-21}
    \label{fig:hUCBQ221}
    \vspace{4ex}
  \end{minipage} 
\end{figure}

Asset quality is measure as the percentage of non-performing loans. The functions in the smart contract automatically accrued the interest on loans and compound these interest to the outstanding value of the loan, therefore in the strict sense, it is difficult to say that there are loans that are not making the scheduled payments because payments are automated. Since the most important risk of the protocol is that the loan value is above the collateral value we can consider as non-performing loans those where the health indicator \ref{eq:health} is below 1, we denote these as undercollateralized borrows (UCB). In table \ref{tbl:camels} we present the previous definition of non-performing loans measures in terms of the outstanding amount of UCB over total borrows, the percentage of accounts UCB and also a ratio of the outstanding amount of UCB over reserves.
The percentage of UCB to total borrows is quite small $0.025\%$ and represent $3.3\%$ of the current value of reserves in June 2021. However, the percentage of accounts categorized as UCB to accounts with positive borrows is $9.2\%$. This indicates that there are many accounts with very small amounts of undercollateralized loans. This is not surprising given the incentive structure of liquidators where there are accounts that are costly to liquidate relatively to the benefit. In addition, close factors restrict full liquidation and are probably responsible for an ever-increasing number of unliquidated accounts with small balances.
\\
In figures \ref{fig:hUCBQ220} through \ref{fig:hUCBQ221} we confirm this by looking at the histogram of UCB outstanding loans in USD. The outstanding of most undercollateralized borrows is less than 3 USD. The extreme values in the distribution indicate that at that particular block time some large positions have not been liquidated. However, they are profitable for a liquidator to step in, for the smaller accounts this is not the case. 

\subsection{Earnings}
\begin{figure}[ht] 
  \label{ fig7} 
  \begin{minipage}[b]{0.5\linewidth}
    \centering
    \includegraphics[width=1\linewidth]{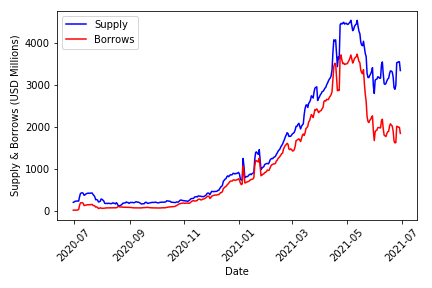} 
    \caption{\scriptsize USDC Total Borrow and Supply in USD Millions Compound}
    \label{fig:usdcAggComp}
    \vspace{4ex}
  \end{minipage}%%
  \begin{minipage}[b]{0.5\linewidth}
    \centering
    \includegraphics[width=1\linewidth]{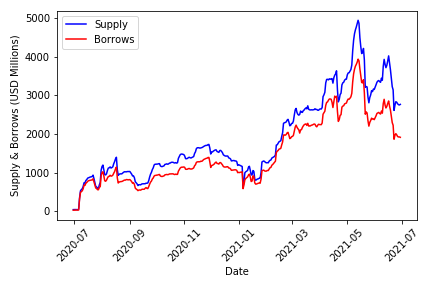} 
    \caption{\scriptsize DAI Total Borrow and Supply in USD Millions Compound} 
    \label{fig:daiAggComp}
    \vspace{4ex}
  \end{minipage} 
  \begin{minipage}[b]{0.5\linewidth}
    \centering
    \includegraphics[width=1\linewidth]{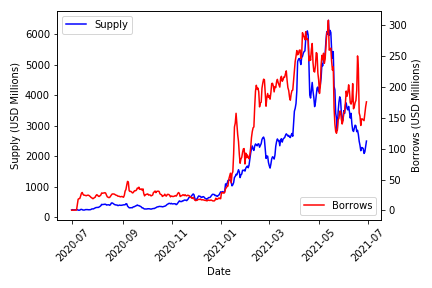} 
    \caption{\scriptsize ETH Total Borrow and Supply in USD Millions Compound}
    \label{fig:ethAggComp}
    \vspace{4ex}
  \end{minipage}%% 
  \begin{minipage}[b]{0.5\linewidth}
    \centering
    \includegraphics[width=1\linewidth]{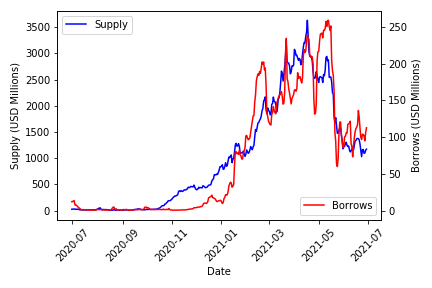} 
    \caption{\scriptsize WBTC Total Borrow and Supply in USD Millions Compound}
    \label{fig:wbtcAggComp}
    \vspace{4ex}
  \end{minipage} 
\end{figure}

For earnings, we use the information used in section \ref{sec:margins} to estimate the ex-post margins: the interest income and expenses in each market in Compound. The operating margin is obtained as the difference between interest income and expenses. The operating margin is express as a ratio to productive assets which are the outstanding loans (total borrows), this is a proxy for return on assets (ROA). Unfortunately, we do not have additional information on expenses from the protocol so we use the operating margin as a proxy for profits to estimate the return on equity (ROE) where the value of the protocol is taken from the market value of the protocol from the current price of the governance token multiplied by the outstanding tokens.
Table \ref{tbl:camels} presents the operational margin derived from the overall interest income and expenses. In the second quarter of 2021 income derived from interest on borrow totalled $176.02$ million dollars, but on the other hand, the accrued interest to suppliers totalled $204.3$ million dollars. This means that the traditional measure of the operating margin is negative ($28.2$ million dollars) and also the ROE and ROA would be negative, $-1.58$ and $-0.8$, respectively\footnote{The average ROE for US banks in the third quarter of 2020 was $5.31$ according to the \hyperlink{https://fred.stlouisfed.org/series/USROE}{FED}.}. This negative operational margin is not surprising given the scale difference between supply and borrow in some markets, in particular, ETH and WBTC, figures \ref{fig:ethAggComp} and \ref{fig:wbtcAggComp}, respectably. An alternative measure of benefits for protocol sponsors is to look at reserves because this is an account that is credited based on a fraction of interest charge to borrowers. Although Compound claims that reserved are exclusive to support the event of systemically undercollateralized accounts, Aave considers that reserves may also be used to support the development of the protocol\footnote{Aave has also developed an insurance mechanism (so call safety module) where owners of the native token, stake their tokens to cover possible losses due to undercollateralization of accounts.}. If we consider that all of it or some part of the reserved can be used to fund the activities of the protocol and benefit its investors then it is interesting to look at the ratio of reserves to equity (where equity is estimated as the market capitalization of the protocols token: COMP). The ratio of reserves over equity has been increasing from $0.134$ in the second quarter of 2020 to $1.529$ a year after.       
\subsection{Sensitivity to market risk}
Market risk is very important to the protocols because of the fluctuating value of the underlying assets used as collateral. Because the collateral in the accounts is valued in ETH, we use daily historical ETH prices in USD from mid-January 2016 to the end of June 2021 to introduce variation in the value of the collateral. We estimate the collateral expected shortfall using price variations over one day and five days (\ref{tbl:camels}). This allows us to estimate the extreme variation in the value of collateral during those time horizons. We find that under extreme market conditions the value of the collateral may fall by $23\%$ over the one-day horizon and $44\%$ over the five-day horizon. This is a stressed scenario because we only consider one source of variation, ETH price, but in reality, a fraction of that collateral may be in stablecoins.

%\section{Covid-19 related stress in decentralized lending platforms}
%The covid pandemic created an oportunity to some of the risk management mechanisms hard coded on these lending protocols. In this section we look at some of the dynamics observed during march 2020.
%discuss some of these metrics during the covid colatility march 2020

%\section{liquidatiors in action}

\section{Conclusions}
\label{sec:conclu}
This paper uses data obtained from the blockchain to provide an overview of financial intermediation and risk in the decentralized lending protocols: Compound and Aave.
These two protocols are two of the most important applications in DeFi deployed in the Ethereum network. The paper provides an overview of how the protocols provide collateralized loans, the participants, risk management practices, historical rates and the number of resources supplied and borrowed. Since the protocols have provided loans like traditional banks we use a traditional approach from the banking literature to measure the cost of financial intermediation using the ex-post margins. The results indicate that the average ex-post margins are around $1\%$ and lower than $0.3\%$ for stablecoins. This is lower than intermediation margins in banking in developed economies and substantially lower than in developing countries. Part of the lower cost of intermediation is because the loans are collateralized where bank loans are not all collateralized. We also used some elements of the CAMELS framework that is standard in banking to characterize the riskiness of the Compound protocol. We observed that the risk of undercollateralization is small compared to current reserves and hence liquidators seem to be well incentivised. It is not clear how to account for the earnings of the protocol, using the traditional approach gives a negative operational margin. The protocol has some other sources of income but for investors and consumers, it is important to provide clear information regarding the benefits of intermediation activities and any additional activity that can be considered as a source of income. A proper understanding and quantification of the earnings are important to avoid any speculative runs on the protocol not based on undercollateralization of accounts (this seems to be under control) but rather because of the expectation that there could be not enough resources to pay interest for all of the suppliers.\\
These alternative financial services are new, they have only been active for a year and a half or less, so there are still many open questions regarding the role these protocols will play in the future of financial intermediations. This paper serves as an open invitation to explore the publicly available data.

 \bibliographystyle{apalike} 
 \bibliography{DexLend}

\clearpage
\newpage

\clearpage
\newpage

\begin{table}
\begin{center}
\begin{tabular}{lccc}
Quarter & \multicolumn{1}{l}{Active} & \multicolumn{1}{l}{Passive} & \multicolumn{1}{l}{Margin} \\
\hline
\textbf{DAI}     &                            &                             &                            \\
II-20   & 1,05                       & 5,31                        & -4,25                      \\
III-20  & 0,95                       & 0,87                        & 0,07                       \\
IV-20   & 2,14                       & 2,00                        & 0,14                       \\
I-21    & 5,71                       & 3,78                        & 1,92                       \\
II-21   & 2,59                       & 2,24                        & 0,35                       \\
\textbf{ETH}     &                            &                             &                            \\
II-20   & 2,57                       & 0,01                        & 2,56                       \\
III-20  & 0,89                       & 0,08                        & 0,81                       \\
IV-20   & 2,21                       & 0,09                        & 2,13                       \\
I-21    & 2,40                       & 0,17                        & 2,22                       \\
II-21   & 1,07                       & 0,08                        & 0,99                       \\
\textbf{USDC}    &                            &                             &                            \\
II-20   & 1,66                       & 0,48                        & 1,19                       \\
III-20  & 2,12                       & 1,30                        & 0,83                       \\
IV-20   & 0,72                       & 0,93                        & -0,21                      \\
I-21    & 5,49                       & 3,33                        & 2,16                       \\
II-21   & 1,20                       & 1,43                        & -0,23                      \\
\textbf{USDT}    &                            &                             &                            \\
II-20   & 0,67                       & 0,44                        & 0,22                       \\
III-20  & 1,89                       & 0,57                        & 1,32                       \\
IV-20   & 0,73                       & 1,27                        & -0,53                      \\
I-21    & 1,28                       & 2,33                        & -1,05                      \\
II-21   & 0,66                       & 1,75                        & -1,09                      \\
\textbf{WBTC}    &                            &                             &                            \\
II-20   & 0,65                       & 0,25                        & 0,40                       \\
III-20  & 2,31                       & 0,45                        & 1,86                       \\
IV-20   & 1,59                       & 0,05                        & 1,54                       \\
I-21    & 1,03                       & 0,23                        & 0,80                       \\
II-21   & 0,48                       & 0,12                        & 0,36  \\
\textbf{All Mkts} &        &         &        \\
II-20   & 0,48                       & 0,45                        & 0,03                       \\
III-20  & 1,25                       & 0,85                        & 0,40                       \\
IV-20   & 1,67                       & 0,93                        & 0,74                       \\
I-21    & 1,98                       & 1,04                        & 0,94                       \\
II-21   & 4,93                       & 2,06                        & 2,87\\               
\hline             
\end{tabular}
\caption{End-of-Quarter ex-post rates and intermediation margin for the most important markets and all of the markets in the Compound protocol.}
\label{tbl:exratesComp}
\end{center}
\end{table}

\begin{table}
\begin{center}
\begin{tabular}{lccc}
Quarter & \multicolumn{1}{l}{Active} & \multicolumn{1}{l}{Passive} & \multicolumn{1}{l}{Margin} \\
\hline
\textbf{DAI}     &                            &                             &                            \\
IV-20   & 0,26                       & 0,18                        & 0,08                       \\
I-21    & 2,09                       & 1,50                        & 0,59                       \\
II-21   & 1,03                       & 0,65                        & 0,38                       \\
\textbf{ETH}     &                            &                             &                            \\
IV-20   & 0,07                       & 0,00                        & 0,07                       \\
I-21    & 3,07                       & 0,04                        & 3,03                       \\
II-21   & 1,90                       & 0,02                        & 1,88                       \\
\textbf{USDC}    &                            &                             &                            \\
IV-20   & 0,26                       & 0,20                        & 0,06                       \\
I-21    & 1,38                       & 1,10                        & 0,28                       \\
II-21   & 1,19                       & 0,72                        & 0,47                       \\
\textbf{USDT}    &                            &                             &                            \\
IV-20   & 0,50                       & 0,38                        & 0,12                       \\
I-21    & 1,37                       & 1,16                        & 0,22                       \\
II-21   & 1,61                       & 1,24                        & 0,37                       \\
\textbf{WBTC}    &                            &                             &                            \\
IV-20   & 0,31                       & 0,01                        & 0,30                       \\
I-21    & 0,22                       & 0,01                        & 0,21                       \\
II-21   & 0,24                       & 0,01                        & 0,23 \\
\hline
\end{tabular}
\caption{End-of-Quarter ex-post rates and intermediation margin for the most important markets in the Aave protocol.}
\label{tbl:exratesAave}
\end{center}
\end{table}

\begin{table}
\begin{center}
\begin{tabular}{l|r}
\hline
$h_{i,t}$ &   Risk Weight  \\
\hline
$h_{i,t} \leq 1$ &  $150\%$   \\
$ 1< h_{i,t} \leq 1.33$ &  $100\%$   \\
$ 1.33< h_{i,t} \leq 2$ &  $50\%$   \\
$ 2< h_{i,t} \leq 10$ &  $20\%$   \\
$ h_{i,t} > 10$ &  $0\%$   \\
\hline
\end{tabular}
\caption{Risk buckets for the borrows in the protocol as a function of the accounts health.}
\label{tbl:riskBuckets}
\end{center}
\end{table}

\begin{sidewaystable}
%\begin{table}
\scriptsize
    \centering
    \begin{tabular}{l|rrrrrrrrr}
Quarter & Borrows                   & RWA                       & UCB                       & Reserves                  & Equity                     & Op. Margin                 & Collateral                & Col. ES 1D                 & Col. ES 5D                 \\
\hline
II-20   & 371.804.526               & 331.392.084               & 552                       & 856.290                   & 637.581.261                & -         2.582.144        & 538.363.233               & 125.801.355                & 240.537.776                \\
III-20  & 1.084.466.118             & 1.349.662.063             & 7.005                     & 2.016.926                 & 442.245.279                & -         2.626.845        & 1.237.483.257             & 284.848.963                & 545.063.745                \\
IV-20   & 1.800.323.709             & 1.990.187.182             & 10.956                    & 3.771.664                 & 619.544.294                & -         4.729.254        & 2.398.649.255             & 544.148.175                & 1.042.341.555              \\
I-21    & 5.556.075.496             & 5.615.725.500             & 614.762                   & 20.871.832                & 1.913.889.788              & -      18.735.076          & 7.673.001.948             & 1.736.022.775              & 3.334.330.256              \\
II-21   & 3.568.787.044             & 3.600.122.326             & 905.031                   & 27.372.147                & 1.790.332.941              & -      28.257.213          & 5.323.799.961             & 1.246.787.198              & 2.394.242.013              \\
\hline
Quarter & Solvency                  & UCB/Reserves              & UCB/Borrows               & UC accounts               & ROE                        & ROA                        & Reserves/Equity          & ColR 1D                    & ColR 5D                    \\
\hline
II-20   & \multicolumn{1}{c}{0,258} & \multicolumn{1}{c}{0,064} & \multicolumn{1}{c}{0,000} & \multicolumn{1}{c}{4,759} & \multicolumn{1}{c}{-0,405} & \multicolumn{1}{c}{-0,694} & \multicolumn{1}{c}{0,134} & \multicolumn{1}{c}{23,367} & \multicolumn{1}{c}{44,679} \\
III-20  & \multicolumn{1}{c}{0,149} & \multicolumn{1}{c}{0,347} & \multicolumn{1}{c}{0,001} & \multicolumn{1}{c}{8,311} & \multicolumn{1}{c}{-0,594} & \multicolumn{1}{c}{-0,242} & \multicolumn{1}{c}{0,456} & \multicolumn{1}{c}{23,018} & \multicolumn{1}{c}{44,046} \\
IV-20   & \multicolumn{1}{c}{0,190} & \multicolumn{1}{c}{0,290} & \multicolumn{1}{c}{0,001} & \multicolumn{1}{c}{7,860} & \multicolumn{1}{c}{-0,763} & \multicolumn{1}{c}{-0,263} & \multicolumn{1}{c}{0,609} & \multicolumn{1}{c}{22,686} & \multicolumn{1}{c}{43,455} \\
I-21    & \multicolumn{1}{c}{0,372} & \multicolumn{1}{c}{2,945} & \multicolumn{1}{c}{0,011} & \multicolumn{1}{c}{8,792} & \multicolumn{1}{c}{-0,979} & \multicolumn{1}{c}{-0,337} & \multicolumn{1}{c}{1,091} & \multicolumn{1}{c}{22,625} & \multicolumn{1}{c}{43,455} \\
II-21   & \multicolumn{1}{c}{0,760} & \multicolumn{1}{c}{3,306} & \multicolumn{1}{c}{0,025} & \multicolumn{1}{c}{9,210} & \multicolumn{1}{c}{-1,578} & \multicolumn{1}{c}{-0,792} & \multicolumn{1}{c}{1,529} & \multicolumn{1}{c}{23,419} & \multicolumn{1}{c}{44,972}\\
\hline
\end{tabular}
\caption{The table presents the quarterly information on user accounts and aggregate protocol variables in USD: stock lo loans/borrows, risk weighted assets (RWA), stock of undercollateralized borrows (UCB), reserves, market value of the protocol (estimated as the market capitalization of the protocols token COMP), operational margin, stock of collateral, Collateral expected shortfall at one and five day horizon. Based on the aggregates we also present some ratios used in the CAMELS framework: solvency ratio, loan loss reserve coverage ratio (measured as the ratio of UCB to reserves), non-performing loans(measured as the ratio between UCB and total borrows), percentage of uncollateralized accounts, ROE, ROA, reserved over borrows and the historical percentage of collateral at risk over the one and five day horizon with respect to total collateral.}
    \label{tbl:camels}
%\end{table}
\end{sidewaystable}

\end{document}